\documentstyle[aps]{revtex}
\begin{document}
\title{Gravitational Waves from Spinning Non-Abelian  \\
Cosmic Strings}
\author{R. J. Slagter}
\address{University of Amsterdam, Physics Department, The Netherlands,\\
SARA, Stichting Academisch Rekencentrum Amsterdam, \\ and \\
ASFYON, Astronomisch Fysisch Onderzoek Nederland\footnote{Electronic address: rjs@asfyon.nl}}
\maketitle
\begin{abstract}
We investigated the  SU(2) Einstein-Yang-Mills system on a time-dependent
non-diagonal cylindrical symmetric spacetime.
From the numerical investigation, wave-like solutions are found,
consistent with the familiar string-like features.
They possess an angle deficit which depends on the the  initial form of the
magnetic component of the Yang-Mills field, i.e., the number of times it crosses
the r-axis. The appearance of a
curvature singularity depends on the ratio of the Planck scale $M_{pl}$
and Yang-Mills couplings constant g, i.e., $\alpha=M_{pl}/g=\frac{1}{\sqrt{4\pi G}g}$
and the initial mass per unit length of the cosmic string.
The solitonlike behavior of gravitational and Yang-Mills waves show significantly
differences from the ones found in the Abelian Einstein-Maxwell system.
The stability of the resulting non-Abelian spinning string is analyzed
using the multiple-scale method. To first order a consistent set of equations is
obtained.

However, a throughout analysis is necessary in order to conclude if the solution is
topological stable and will have magnetic charge.
\end{abstract}

\section{Introduction}

Taking the Yang-Mills model as the matter part of the equations of Einstein,
one finds a rich spectrum of stationary regular and non-regular solutions.
This in contrast
with the vacuum solution and Einstein-Maxwell(EM) solutions, where the Schwarzschild
respectively the Reissner-Nordstr{\"o}m(RN) solution are the only static black hole
solutions. In the Einstein-Yang-Mills(EYM) theory one finds not only the embedding of the
Abelian RN black hole, but in addition there are genuine non-Abelian coloured black
holes. Their co-existence gives rise to the violation of the no-hair conjecture,
since they carry the same (magnetic) charge. For an overview, see Volkov
and Gal'tsov~\cite{volk99}.
After the discovery by Bartnik and McKinnon(BMK)~\cite{bart88}
of these non-Abelian particle-like and later the non-RN (hairy) black hole solutions
of the SU(2) EYM theory~\cite{smol96}, throughout instability
investigations where done by several authors~\cite{strau190,lav95,strau290,zhou91,breit97}.
It turns out that under spherically symmetric perturbations the BMK solution
is unstable. The nth BMK solution has in fact 2n unstable modes,
comparable with the flat space-time 'sphaleron' solution. In the static case,
it was recently conjectured~\cite{mav98} that there are hairy black holes
in the SU(N) EYM-theories with topological instabilities.
In order to analyze the stability of the solutions, one usually linearized
the field equations~\cite{strau190,strau290}.
We conjectured that the conventional linear analysis is in fact inadequate
to be applied to the situation where a singularity is formed~\cite{Slag398}.
One better can apply the so-called multiple-scale or two-timing method,
developed by Choquet-Bruhat~\cite{choq169,choq277,taub80}. This method
is particularly useful for constructing uniformly valid approximations to
solutions of perturbation problems~\cite{slag191}.
At the threshold of black hole
formation it is found numerically~\cite{chop193,chop296,garf97,bizon96}
that there is a critical parameter
whose value separates solutions containing black holes
from those which do not. For a critical value one observes self-similarity:
it produces itself (echoes) on progressively finer scales (Choptuik-scaling).
It is evident that the collapsing ball of field energy will produce
gravitational waves, which will be coupled to the  YM-field perturbations.
Due to the accumulation of echoes, the curvature diverges.
Close related to this critical behavior is the comparable irregularities which
can be found in the Einstein-Skyrme(ES) model and the Einstein-Yang-Mills-Higgs(EYMH)
model. In the latter model, space-like hypersurfaces develop for a critical
value of $\alpha =M_W/gM_{pl}$ two distinct regions separated by a long throat,
with $M_W$ the YM scale and $M_{pl}=1/\sqrt{4\pi G}$.

The solutions mentioned above need not be spherically symmetric. Since the
pioneering work of Stachel~\cite{Sta66}, we know that stationary
cylindrical symmetric rotating models possess intriguing features.
First, the evolution of the 'classical' cylindrical gravitational wave-solution
in the vacuum situation can be described in terms of two effects: the usual
cylindrical reflections and a rotation of the polarization vector of the waves,
an effect comparable with the Faraday rotation~\cite{Piran85}.
Secondly, the time-dependent solutions can be related to the cosmic-string
solution interacting with gravitational waves~\cite{Xan186}.
Usually the cosmic string solution, with its famous conical structure
,is found in the U(1)-gauge Einstein-Higgs system.
It is remarkable that the cylindrical rotating vacuum solution possesses the
same conical structure. Moreover, it is asymptotically flat and free
of singularities.
The physical interpretation can nicely be formulated using the Belinsky-Zakharov
method of integrating the Einstein equations with the help of the Ernst
potentials~\cite{Eco88}. Using the C-energy description, one can conclude that the
solution represents solitonic gravitational waves interacting with a
straight-line cosmic string, where the asymptotic angle deficit depends
on the C-energy of the gravitational waves.
When  a Maxwell field is incorporated, the physical interpretation of the solutions
is not easily formulated. It is hard to believe that in the electrovac solution
the electro-magnetic waves do not contribute to the angle deficit~\cite{Xan287}.
Moreover, the mass per unit length of the cosmic string, which is a constant
in the Xanthopoulos solution, will be determined by the energy-momentum tensor
of the non-vacuum situation. In stead of 'electrifying' or 'magnetifying' the
vacuum solutions via the Ernst formulation, one can better consider the coupled
system of Einstein equations and matter field equations.

Here we consider the Einstein-Yang-Mills system on a cylindrical symmetric
rotating space-time. This situation con formally be obtained from the
axially symmetric situation by complex substitution of $t\to iz, z\to it$~\cite{kram80}.
In the non-rotating cylindrically symmetric situation, it was numerically found
~\cite{slag299} that the singular behavior at finite distance of the core as found
in the supermassive situation, will be pushed to infinity for some critical values
of the gauge coupling constant and one of the YM components.

We will use a (2+1)+1 reduction scheme~\cite{Mae80}
in order to obtain the suitable parameterization for the Yang-Mills potential.
In fact one replaces the YM field by a 2D Abelian one-form, a 2D Kaluza-Klein
two-form, a 2D matrix-valued scalar and a 3D Higgs field~\cite{gal98}.

In the Abelian counterpart model it is known that the space-time around
(spinning) gauge strings could have, besides the usual angle deficit feature,
exotic properties, such as the violation of causality~\cite{Jen92,Sol92},
time-delay effects~\cite{Har88}, helical structure
of time~\cite{Des92,Let95,Maz86} and frame-dragging~\cite{Bon91}.
A class of approximate solutions of the coupled Einstein-scalar-gauge field equations
on the ($r,z$)-plane of an axially symmetric space-time was found~\cite{slag695}.
So a question addressed in this paper will be whether the string features will be
maintained in the EYM system.
In the general $(r,z)$-dependent situation, a nice overview was given by
Islam~\cite{Islam85}. One can proof via the Ernst formulation that for any solution
of the stationary
Einstein equations one can generate a corresponding solution of the EM equations.
In fact, one introduces two complex potentials, for which one finds two second-order
elliptic differential equations.
Starting with the Kerr solution, it is possible to use this correspondence to find
a solution of the EM equations leading to the Kerr-Newman solution.
This 'Ernst-route' fails in the EYM system, simply by the fact that one
of the metric components does not decouple from the other ones.
It is also known that the electrovac staticity theorem~\cite{heus93} does not generalize to the EYM
system. In fact, spinning EYM systems must be electrically charged. So another question
addressed in this paper is whether electrically and magnetically charged solution will exist
in the EYM system.
Finally, this is not the complete story: the investigations can be extended to the
model where the interior solution is  properly matched onto the exterior
solution of the string~\cite{slag496}. This is currently under study by the author.

The plan of this paper is as follows. In Sec. II we derive the field equation
on an cylindrical symmetric space time. In Sec. III we present the numerical solution.
In section IV we present a approximate solution using the multiple-scale method and
in Sec. V we summarize and analyse our results.

\section{The field equations of the Einstein Yang-Mills system}

Consider the Lagrangian of the SU(2) EYM system
\begin{equation}
S=\int d^4x\sqrt{-g}\Bigl[\frac{{\cal R}}{16\pi G}-
\frac{1}{4}{\cal F}_{\mu\nu}^a{\cal F}^{\mu\nu a}\Bigr],
\end{equation}
with the YM field strength
\begin{equation}
{\cal F}_{\mu\nu}^a =\partial_\mu A_\nu^a-\partial_\nu
A_\mu^a+g\epsilon^{abc}A_\mu^bA_\nu^c,
\end{equation}
g the gauge coupling constant, G Newton's constant, $A_\mu^a$ the gauge potential,
and ${\cal R}$ the curvature scalar. The field equations then become
\begin{equation}
G_{\mu\nu}=8\pi G {\cal T}_{\mu\nu},
\end{equation}
\begin{equation}
{\cal D}_\mu {\cal F}^{\mu\nu a}=0,
\end{equation}
with ${\cal T}$ the energy-momentum tensor
\begin{equation}
{\cal T}_{\mu\nu}={\cal F}_{\mu\lambda}^a{\cal F}_\nu^{\lambda a}
-\frac{1}{4}g_{\mu\nu}{\cal F}_{\alpha\beta}^a {\cal F}^{\alpha\beta a},
\end{equation}
and ${\cal D}$ the gauge-covariant derivative,
\begin{equation}
{\cal D}_\alpha{\cal F}
_{\mu\nu}^a\equiv \nabla_\alpha{\cal F}_{\mu\nu}^a+g\epsilon^{abc}A_\alpha^b{\cal F}
_{\mu\nu}^c.
\end{equation}
Let us consider the cylindrical space time
\begin{equation}
ds^2=-\Omega^2(dt^2-dr^2)+f(dz+\omega d\varphi)^2+\frac{r^2}{f}d\varphi ^2,
\end{equation}
where $\Omega, f$ and $\omega $ are functions of t and r.
The problem is, which ansatz one needs for the YM field.
It is often argued that Einsteins gravity on a three dimensional
space time shows some similarity with Chern Simons theory with the Poincar\'e
group being the underlying gauge group. So when trying to find the dimensional
reduction for the stationary axially symmetric EYM system, one can be guided by
the results found by using the (2+1)+1 reduction scheme suggested by Maeda, Sasaki,
Nakamura and Miyama~\cite{Mae80}, and later worked out by Gal'tsov~\cite{gal98}.
One writes in general
\begin{equation}
ds^2=-e^\psi (dt+v_idx^i)^2+e^{2\phi-\psi}(d\varphi +k_adx^a)^2+g_{ab}dx^adx^b,
\end{equation}
where $v_idx^i=\omega (d\varphi +k_adx^a)+\nu_adx^a, k_a$ an one-form, generating
a field strength  $\kappa_{ab}=\partial_ak_b-\partial_bk_a$. The one-form $\nu_a$ give rise to the
2D field strength $\omega_{ab}=\partial_a\nu_b-\partial_b\nu_a+\omega\kappa_{ab}$(a,b=1,2).
The YM potential is then written as
\begin{equation}
A_\mu dx^\mu ={\bf a}_adx^a +{\bf \Phi}(dt+\nu_idx^i)+{\bf \Psi}(d\varphi +k_adx^a)
\end{equation}
Here ${\bf a}$ is the dynamical part of the YM field, parameterized by a 2D complex
Abelian one-form, ${\bf \Phi}$ a 3D Higgs field and ${\bf \Psi}$ a matrix-valued
scalar. In our simplified case of metric (7), we have $\nu_1=-\omega k_1, k_2=\nu_2 =0$ and
the YM potential can be as
\begin{equation}
A_\mu ={\bf \Phi}(dz+\omega d\varphi )+{\bf \Psi}(d\varphi -k_1dt) +{\bf a}_adx^a
\end{equation}
The YM-part of the Lagrangian can then be written as

\begin{eqnarray}
{\cal L}_{YM}=\Omega^2r\Bigl[\Bigl\{f_{ab}f^{ab}+\frac{2f}{r^2}(D_a{\bf\Psi} +{\bf\Phi}
\partial_a\omega -a_a^{'})(D^a{\bf\Psi}+{\bf\Phi}\partial^a\omega
 -a^{' a}) \Bigr\} \cr +\frac{2}{f}\Bigl\{D_a{\bf\Phi}D^a{\bf\Phi}
 +\frac{f}{r^2}({\bf\Phi}^{'}+g[{\bf\Phi},{\bf\Psi}])^2\Bigr\}\Bigl],
\end{eqnarray}
with
\begin{equation}
D_a{\bf\Psi}=\partial_a{\bf\Psi}+g[a_a,{\bf\Psi}]-k_a{\bf\Psi}^{'},
\end{equation}
$f_{ab}$ the 2D field strength
\begin{equation}
f_{ab}=\partial_a a_b-\partial_b a_a +g[a_a,a_b]+a_a^{'}k_b-a_b^{'}k_a+{\bf \Phi}\omega_{ab}+
{\bf \Psi}k_{ab},
\end{equation}
and a prime denoting the partial derivative with respect to $\varphi$.

Let us consider the following parameterization
\begin{eqnarray}
a_1=A_2(t,r)\tau_\varphi ,\quad &&a_2=A_1(t,r)\tau_\varphi,\quad
{\bf\Psi}=W_1(t,r)\tau_r +(W_2(t,r)-1)\tau_z,\cr
&& {\bf\Phi}=\Phi_1(t,r)\tau_r+\Phi_2(t,r)\tau_z,
\end{eqnarray}
with $\tau_i$ the usual set of orthonormal vectors.
From the condition $T_{tt}-T_{rr}=0$, we obtain
\begin{equation}
\partial_t A_1=\partial_r A_2,\quad  W_2=\frac{\Phi_2 W_1}{\Phi_1}+\frac{g-1}{g}.
\end{equation}

From the Einstein equations and the YM equations we obtain the set of
partial differential equations for $\Omega , f , \omega , \Phi_1 , \Phi_2$
and $W_1$.
It follows  from a combination of the YM equations, that $A_1$
and $A_2$ can be expressed in $\Phi_1$ and $\Phi_2$:
\begin{eqnarray}
A_1 =\frac{\Phi_1\partial_r\Phi_2 -\Phi_2\partial_r\Phi_1}{g(\Phi_1^2+\Phi_2^2)},
\qquad A_2 =\frac{\Phi_1\partial_t\Phi_2-\Phi_2\partial_t\Phi_1}{g(\Phi_1^2+\Phi_2^2)}
\end{eqnarray}
For the gauge $\partial_t A_2=\partial_r A_1$
and hence $\partial_t^2 A_i-\partial_r^2 A_i =0$, we obtain from the Einstein equations Eq.(3)
\begin{eqnarray}
\partial_t^2\Omega -\partial_r^2\Omega +\frac{f^2\Omega}{4r^2}\Bigl[(\partial_t\omega )^2
-(\partial_r\omega)^2\Bigr]+\frac{1}{\Omega}\Bigl[(\partial_r\Omega )^2
-(\partial_t\Omega)^2\Bigr]+
\frac{\Omega}{4f^2}\Bigl[(\partial_t f)^2-(\partial_r f)^2\Bigr]
+ \frac{\Omega}{2fr}\partial_r f=0,
\end{eqnarray}
\begin{eqnarray}
\partial_t^2 f -\partial_r^2 f +\frac{1}{f}\Bigr[(\partial_r f)^2-(\partial_t f)^2
\Bigr]+\frac{f^3}{r^2}\Bigl[(\partial_r\omega)^2-(\partial_t\omega)^2\Bigr]
-\frac{1}{r}\partial_r f = \cr \frac{8\pi G}{r^2\Phi_1^4}\Bigl[(\Phi_1^4r^2-f^2\Phi_2^2W^2)(
(\partial_r\Phi_1)^2-(\partial_t\Phi_1)^2) +(\Phi_1^4r^2-f^2\Phi_1^2W^2)((\partial_r\Phi_2)^2
-(\partial_t\Phi_2)^2) \cr +f^2\Phi_1^2(\Phi_1^2+\Phi_2^2)\Bigl((\partial_tW)^2-(\partial_rW)^2
+\Phi_1^2\Bigl((\partial_t\omega)^2-(\partial_r\omega)^2\Bigr)\Bigr) \cr
+2f^2\Phi_1^3(\Phi_1^2+\Phi_2^2)(\partial_t\omega\partial_tW-\partial_r\omega\partial_rW)
+2f^2\Phi_2^2\Phi_1W(\partial_rW\partial_r\Phi_1-\partial_tW\partial_t\Phi_1)\cr
+2f^2\Phi_1^3\Phi_2W(\partial_t\Phi_2\partial_t\omega -\partial_r\Phi_2\partial_r\omega )
+2f^2\Phi_1^2\Phi_2W(\partial_tW\partial_t\Phi_2-\partial_rW\partial_r\Phi_2)\cr
+2f^2\Phi_1\Phi_2W^2(\partial_r\Phi_2\partial_r\Phi_1-\partial_t\Phi_2\partial_t\Phi_1)
+2f^2\Phi_2^2\Phi_1^2W(\partial_r\Phi_1\partial_r\omega-\partial_t\Phi_1\partial_t\omega )\cr
+g\Phi_1^2(\Phi_1^2r^2-f^2W^2)\Bigl(2A_1(\Phi_2\partial_r\Phi_1-\Phi_1\partial_r\Phi_2)
+2A_2(\Phi_1\partial_t\Phi_2-\Phi_2\partial_t\Phi_1)+g(\Phi_1^2+\Phi_2^2)(A_1^2-A_2^2)\Bigl)\Bigl]
\end{eqnarray}
\begin{eqnarray}
\partial_t^2\omega -\partial_r^2\omega +\frac{2}{f}(\partial_tf\partial_t\omega
-\partial_rf\partial_r\omega )+\frac{1}{r}\partial_r\omega
=\frac{16\pi G}{f\Phi_1^2}\Bigl[W\Phi_1 ((\partial_r\Phi_2)^2-(\partial_t\Phi_2)^2)\cr
+\Phi_2W(\partial_t\Phi_2\partial_t\Phi_1-\partial_r\Phi_2\partial_r\Phi_1)
+\Phi_2\Phi_1^2(\partial_r\Phi_2\partial_r\omega -\partial_t\Phi_2\partial_t\omega )
+\Phi_2\Phi_1(\partial_r\Phi_2\partial_rW-\partial_t\Phi_2\partial_tW) \cr
+\Phi_1^2(\partial_r\Phi_1\partial_rW-\partial_t\Phi_1\partial_tW)
+\Phi_1^3(\partial_r\Phi_1\partial_r\omega-\partial_t\Phi_1\partial_t\omega ) \cr
+g^2\Phi_1W(\Phi_1^2+\Phi_2^2)(A_1^2-A_2^2)
+2g\Phi_1^2W(A_2\partial_t\Phi_2-A_1\partial_r\Phi_2)+2g\Phi_1\Phi_2W(A_1\partial_r\Phi_1
-A_2\partial_t\Phi_1)\Bigr],
\end{eqnarray}
and from the YM equations Eq. (4)
\begin{eqnarray}
\partial_t^2\Phi_1-\partial_r^2\Phi_1-\frac{1}{r}\partial_r\Phi_1+\frac{f^2\Phi_1}{r^2}
\Bigl((\partial_r\omega)^2-(\partial_t\omega)^2\Bigl)
+\frac{1}{f}(\partial_rf\partial_r\Phi_1-\partial_tf\partial_t\Phi_1) \cr
+\frac{f^2}{r^2}(\partial_r\omega\partial_rW-\partial_t\omega\partial_tW)
+\frac{gf^2\Phi_2W}{\Phi_1r^2}(A_1\partial_r\omega - A_2\partial_t\omega )\cr
+\frac{g\Phi_2}{f}(A_1\partial_rf-A_2\partial_tf)
+2g(A_2\partial_t\Phi_2-A_1\partial_r\Phi_2)-\frac{gA_1\Phi_2}{r}+g^2\Phi_1(A_1^2-A_2^2)=0
\end{eqnarray}
\begin{eqnarray}
\partial_t^2\Phi_2-\partial_r^2\Phi_2 -\frac{1}{r}\partial_r\Phi_2+\frac{f^2\Phi_2}{r^2}
\Bigl((\partial_r\omega)^2-(\partial_t\omega)^2\Bigr)
+\frac{1}{f}(\partial_rf\partial_r\Phi_2-\partial_tf\partial_t\Phi_2)
+\frac{f^2\Phi_2}{r^2\Phi_1}(\partial_r\omega\partial_rW-\partial_t\omega\partial_tW)\cr
+\frac{f^2\Phi_2W}{r^2\Phi_1^2}(\partial_t\omega\partial_t\Phi_1-\partial_r\omega\partial_r\Phi_1)
+\frac{f^2W}{r^2\Phi_1}(\partial_r\omega\partial_r\Phi_2-\partial_t\omega\partial_t\Phi_2)
+2g(A_1\partial_r\Phi_1-A_2\partial_t\Phi_1) \cr +\frac{g\Phi_1}{f}(A_2\partial_tf-A_1\partial_rf)
+\frac{gf^2W}{r^2}(A_2\partial_t\omega-A_1\partial_r\omega)+g^2\Phi_2(A_1^2-A_2^2)
+\frac{gA_1\Phi_1}{r}=0
\end{eqnarray}
\begin{eqnarray}
\partial_t^2 W_1-\partial_r^2 W_1 +\frac{1}{r}\partial_rW
+\frac{1}{f}(\partial_tf\partial_tW-\partial_rf\partial_rW)
+\frac{\Phi_1}{f}(\partial_rf\partial_r\omega -\partial_tf\partial_t\omega ) \cr
+\partial_t\omega\partial_t\Phi_1-\partial_r\omega\partial_r\Phi_1
+\frac{2g\Phi_2}{\Phi_1}(A_2\partial_tW-A_1\partial_rW)
+g\Phi_2(A_2\partial_t\omega-A_1\partial_r\omega) \cr
+\frac{2gW\Phi_2}{\Phi_1^2}(A_1\partial_r\Phi_1-A_2\partial_t\Phi_1)+
\frac{2gW}{\Phi_1}(A_2\partial_t\Phi_2-A_1\partial_r\Phi_2)  +\frac{gW\Phi_2}{f\Phi_1}
(A_2\partial_tf-A_1\partial_rf)\cr +g^2W(A_1^2-A_2^2)+\frac{gA_1\Phi_2W}{r\Phi_1}
-\frac{16\pi G}{f}\Bigl[W\Bigl((\partial_t\Phi_2)^2-(\partial_r\Phi_2)^2\Bigr)\cr
+\Phi_1(\partial_t\Phi_1\partial_tW-\partial_r\Phi_1\partial_rW)
+\Phi_1^2(\partial_t\omega\partial_t\Phi_1-\partial_r\omega\partial_r\Phi_1) \cr
+\Phi_2(\partial_t\Phi_2\partial_tW-\partial_r\Phi_2\partial_rW)
+\frac{W\Phi_2}{\Phi_1}(\partial_r\Phi_2\partial_r\Phi_1-\partial_t\Phi_2\partial_t\Phi_1)
+\Phi_1\Phi_2(\partial_t\Phi_2\partial_t\omega-\partial_r\Phi_2\partial_r\omega) \cr
+2gW\Phi_1(A_1\partial_r\Phi_2-A_2\partial_t\Phi_2)
+2gW\Phi_2(A_2\partial_t\Phi_1-A_1\partial_r\Phi_1)+g^2\Phi_1^2W(\Phi_1^2+\Phi_2^2)(A_2^2-A_1^2)\Bigr]=0.
\end{eqnarray}

\section{Numerical solution}
In the vacuum situation, the behavior of the cylindrical wave solution is
well established~\cite{Bon294,Tom89}. One can have reflection of incoming
into outgoing waves with the same polarization. Further, there is
a non-linear effect describing the interaction between the two different
polarization $+$ and $\times$ modes. If an
outgoing cylindrical wave is linear polarized, its polarization vector
rotates as it propagates. This effect  is often called 'Faraday' rotation and is given by
\begin{equation}
\tan\theta_A=\frac{A_x}{A_+},\quad \tan\theta_B=\frac{B_x}{B_+}
\end{equation}
with
\begin{equation}
A_{+,x}=\frac{1}{f}(\partial_tf\pm\partial_rf),\quad B_{+,x}=\frac{f}{r}(\partial_t\omega
\pm\partial_r\omega)
\end{equation}

A wave emitted from the axis of symmetry behaves like a shock
front and is given by
\begin{equation}
f=w(a^2+1)/(a^2+w^2),\quad \omega=ra(1-w^2)/w(a^2+1),
\quad \Omega^2 f=b\sqrt{r}/\sqrt{t^2-r^2},
\end{equation}
with $w=(t-\sqrt{t^2-r^2})/r$ and a and b arbitrary constants.
One can generate a two-soliton solution and it is found that near the axis
$\log (\Omega^2 f)$, which also describes the 'C-energy',
does not approach zero. So some matter field
must act as a possible source of the two-soliton field. This curious
conclusion can also be formulated in context of the cosmic string with mass
unit length $m=p/(2p-2)$, with p a constant originating from the static solution
$f=r^{2p}, \Omega =ar^{p^2-p}$, which on his turn is related to the
angle deficit of the string, given by
\begin{equation}
\delta\phi=2\pi(1-\frac{1}{\sqrt{\Omega^2f}})
\end{equation}
When a 'Rosen'-pulse $\ln f =\int_0^{t-r}f(\beta)d\beta / \sqrt{(t-\beta)^2-r^2}$ is emitted,
$\log (\Omega^2 f)$ changes by an amount $-k^2$, where k depends solely on $f(\beta )$.
So after the pulse, one has a different string, which let Marder~\cite{mar58}
to conclude that the mass per unit length has decreased by an amount
$\sim k^2$. Because $\log (\Omega^2f)$ is related to the 'C'-energy, this confirms that
the waves carry away energy and are evidently generated by a source on the
axis. So a continuous change in the parameter p  may produce suddenly
and inexplicable change in physical meaning.

Further, it was emphasized in the vacuum case that there is a reflection of the outgoing
wave into an ingoing wave and visa versa.

From the Einstein-Maxwell case~\cite{Xan186,Xan287} it is known that there exist
stable open cosmic strings coupled with gravity and electromagnetism.
When gravitational and electromagnetic waves are generated,the angle deficit
is increased asymptotically and near the axis independent of the coupling
of the Maxwell field to gravity.
Further, it was emphasized that the electromagnetic waves do contribute
to the rotation of the spacetime by effecting the 'dragging' of the Killing field
$\frac{\partial}{\partial z}$, but not to the angle deficit.

In our EYM case we solved Eq.(17)-(22) for  the initial values
\begin{eqnarray}
\Omega(0,r)=C_1r^{p^2-p},\quad f(0,r)=r^{2p}, \quad \Phi_1(0,r)=1-e^{-C_2r},
\quad \Phi_2(0,r)=e^{-C_3r}-1
\end{eqnarray}
For  $W_1$ we will take as initial value a node number 1
or node number 2 function, i.e., $W_1(0,r)=-\tanh(r-1)$, or $W_1 (0,r)=1-e^{-C_4r^2}$.
Here are $C_i$ some constants.

We used a 100x100 grid and applied the method of lines with cubic Hermite polynomials.
The roundoff error remained below 0.001.
In figure 1 we took  $\omega(0,r)=\frac{3.5r}{r^2+1}$ and for $W_1$ the
node number 2 function. We plotted respectively: $\Omega$, f, $\omega$, $\Phi_1$,
$W_1$, $\Phi_2$, the angle deficit, the gravitational puls wave $\Omega^2f$, $g_{\varphi\varphi}$,
and $W_2$.
In figure 2 we took a different initial $\Omega$ (p=0.1 in stead of 0.25).
We obtain a solitonlike solution. The behavior depends  on the initial mass
per unit length of the string, just as in the vacuum case.
In figure 3 we took the node number 1 function for $W_1$. We observe that the angle deficit
changes significantly as well as the gravitational outgoing pulse. So the Yang-Mills waves
has an impact on the gravitational waves. In figure 4 we took p=0.1 in stead of
p=0.25. We see that the behavior of the solution changes significantly: the angle deficit
close to the r-axis increases, while the gravitational puls 'hangs' close to the r-axes.

In figure 5 we plotted a long-time run for a different initial $\omega$: an outgoing wave
is reflected into an ingoing wave.

An important question is if the solution admits gauge-charges.
The gauge-charges $Q_E,Q_M$ (electric and magnetic) are given by

\begin{eqnarray}
Q_E=\frac{1}{4\pi}\oint_S\vert ^*F\vert,\quad Q_M=\frac{1}{4\pi}\oint_S\vert F\vert,
\end{eqnarray}
with F the YM field strength, $^*F$ the dual tensor and the integration over
a two-sphere at spatial infinity.
In our situation, we obtain for the charges
\begin{eqnarray}
Q_E=\int r\Bigl\{\Bigl[\partial_t(W_1+\omega\Phi_1)+
gA_2(W_2-1+\omega\Phi_2)+A_2\Bigr]\tau_r +\Bigl[\partial_t(W_2+\omega\Phi_2)
-gA_2(W_1+\omega\Phi_1)\Bigr]\tau_z\Bigr\}dr
\end{eqnarray}
\begin{eqnarray}
Q_M=\int\frac{r}{\Omega^2} \Bigl\{\Bigl[\frac{1}{f}(\partial_r\Phi_1+gA_1\Phi_2
-\frac{f\omega}{r^2}(\Phi_1\partial_r\omega+\partial_rW_1+A_1+gA_1(W_2-1))\tau_r\Bigr] \cr
+\Bigl[\frac{1}{f}(\partial_r\Phi_2-gA_1\Phi_1)-\frac{f\omega}{r^2}(\partial_rW_2+\Phi_2\partial_r\omega-gA_1W_1)\Bigr]\tau_z)\Bigr\}dr,
\end{eqnarray}
where we used Eq. (15). So in general the solution will possess gauge charges.

\section{An approximate wave solution to first order}

In the appendix we presented the multiple-scale method for the EYM system.
On the metric Eq. (7) we have for the tetrad component:
$l_\mu =(-1,1,0,0)$ and
$l^\mu =\frac{1}{\Omega^2}(1,1,0,0)$, which implies that the divergence of the
null congruence $l^\mu$ becomes
\begin{equation}
\nabla_\mu l^\mu =\frac{1}{\Omega^2 r},
\end{equation}
as it should be.
From Eq. (A10) we obtain $B_0^a=-B_1^a$, or, in the notation of Eq. (14),
$\dot A_1^{(1)}=-\dot A_2^{(1)}$.
From Eq. (A8) we obtain
\begin{eqnarray}
\dot h_{01}=-\frac{1}{2}(\dot h_{00}+\dot h_{11}),\quad
\dot h_{12}=-\dot h_{02},\quad \dot h_{03}=-\dot h_{13},
\quad \dot h_{23}=-\frac{\bar g^{22}\dot h_{22}+\bar g^{33}\dot h_{33}}{2\bar g^{23}}.
\end{eqnarray}
First of all we have just as Eq. (15),
\begin{eqnarray}
\partial_t\bar A_1=\partial_r\bar A_2,\quad \bar W_2=\frac{\bar \Phi_2\bar W_1}{\bar \Phi_1}
+1-\frac{1}{g}
\end{eqnarray}
For simplicity we will take $\Phi_2=0$ and try to find an approximate wave solution to second order.
From the propagation equations Eq. (A14) and (A18) we obtain
\begin{eqnarray}
\bar\Phi_1\partial_\varphi\dot \omega^{(1)}+\partial_\varphi\dot W_1^{(1)}=0,
\end{eqnarray}
\begin{eqnarray}
\partial_\varphi\dot W_2^{(1)}=0.
\end{eqnarray}
So we will take $\dot\omega^{(1)}$ and $\dot W_2^{(1)}$ independent of $\varphi$.
Further,
\begin{eqnarray}
\partial_t\dot A_1^{(1)}+\partial_r\dot A_1^{(1)}
=\ddot A_1^{(2)}+\ddot A_2^{(2)}-\frac{g\bar f\bar\Omega^2}{r^2}\bar W_1\dot W_2^{(1)},
\end{eqnarray}

\begin{eqnarray}
\partial_t\dot\Phi_1^{(1)}+\partial_r\dot\Phi_1^{(1)}=-\frac{\dot\Phi_1^{(1)}}{2r}
+\frac{\dot f^{(1)}}{2\bar f}(\partial_t\bar\Phi_1 +\partial_r\bar\Phi_1 )
+\frac{\dot\Phi_1^{(1)}}{2\bar f}(\partial_t\bar f +\partial_r\bar f )
\cr +\frac{\bar f^2}{r^2}(\dot\omega^{(1)}\bar\Phi_1+\frac{1}{2}\dot W_1^{(1)})
(\partial_t\bar\omega +\partial_r\bar\omega )
+\frac{\bar f^2\dot\omega^{(1)}}{2r^2}(\partial_t\bar W_1+\partial_r\bar W_1),
\end{eqnarray}
\begin{eqnarray}
\partial_t\dot W_2^{(1)}+\partial_r\dot W_2^{(1)}=-\frac{1}{2\bar f}\dot W_2^{(1)}
(\partial_t\bar f+\partial_r\bar f)+\frac{1}{2r}\dot W_2^{(1)}+g(\bar A_1+\bar A_2)
\Bigl(\dot W_1^{(1)}+\frac{\bar\Phi_1\dot\omega^{(1)}}{2}+\frac{\dot f^{(1)}\bar W_1}{2\bar f}
\Bigr)
\end{eqnarray}

\begin{eqnarray}
\partial_t\dot W_1^{(1)}+\partial_r\dot W_1^{(1)}=\frac{1}{2r}\dot W_1^{(1)}
-\frac{1}{2\bar f} (\dot W_1^{(1)} +\bar\Phi_1\dot\omega^{(1)})
(\partial_t\bar f+\partial_r\bar f)
-\bar\Phi_1(\partial_t\dot\omega^{(1)}+\partial_r\dot\omega^{(1)})-
\frac{\dot f^{(1)}}{2\bar f}(\partial_r\bar W_1+\partial_t\bar W_1)\cr
-\frac{1}{2}(\partial_t\bar\omega+\partial_r\bar\omega)(\dot\Phi_1^{(1)}+\frac{\bar\Phi_1
\dot f^{(1)}}{\bar f})-\frac{\dot\omega^{(1)}}{2}(\partial_t\bar\Phi_1+\partial_r\bar\Phi_1)
+\frac{\bar\Phi_1\dot\omega^{(1)}}{2r}-g\dot W_2^{(1)}(\bar A_1+\bar A_2)
\end{eqnarray}
\begin{eqnarray}
\partial_t\dot\Omega^{(1)}+\partial_r\dot\Omega^{(1)}=-\frac{\bar\Omega\dot f^{(1)}}{4\bar f^2}
(\partial_t\bar f+\partial_r\bar f)+\frac{\dot\Omega^{(1)}}{\bar\Omega}
(\partial_t\bar\Omega +\partial_r\bar\Omega)+\frac{\bar\Omega\dot f^{(1)}}{4r\bar f}
+\frac{1}{4\bar\Omega}(\ddot k_{00}+\ddot k_{11}+2\ddot k_{01})\cr
-\frac{8\pi G \bar\Omega \bar f}{r^2}\Bigl[\frac{\dot\Phi_1^{(1)}}{\bar f^2}
(\partial_t\bar\Phi_1+\partial_r\bar\Phi_1)
+(\dot\omega^{(1)}\bar\Phi_1+\dot W_1^{(1)})\Bigl(\partial_t\bar W_1+\partial_r\bar W_1
+\bar\Phi_1(\partial_t\bar\omega
+\partial_r\bar\omega)\Bigr)-g\dot W_2^{(1)}\bar W_1(\bar A_1+\bar A_2)\Bigr]
\end{eqnarray}

\begin{eqnarray}
\partial_t\dot f^{(1)}+\partial_r\dot f^{(1)}=\frac{\dot f^{(1)}}{\bar f}(\partial_t \bar f
+\partial_r\bar f)-\frac{\dot f^{(1)}}{2r}+\frac{\bar f^3(\partial_t\bar\omega +
\partial_r\bar\omega )}{r^2}\dot\omega^{(1)}-\frac{8\pi G\bar f^2}{r^2}\Bigl((
\dot\omega^{(1)}\bar\Phi_1+\dot W_1^{(1)})(\partial_t\bar W_1+\partial_r\bar W_1) \cr
+(\bar\Phi_1^2\dot\omega^{(1)}+\bar\Phi_1\dot W_1^{(1)})(\partial_t\bar\omega
+\partial_r\bar\omega)
+\frac{3r^2\dot\Phi_1^{(1)}}{\bar f^2}(\partial_t\bar\Phi_1+\partial_r\bar\Phi_1)\Bigr),
\end{eqnarray}
and
\begin{eqnarray}
\partial_t\dot\omega^{(1)}+\partial_r\dot\omega^{(1)}=\frac{1}{2r}\dot\omega^{(1)}
-\frac{\dot f^{(1)}}{\bar f}(\partial_t\bar\omega +\partial_r\bar\omega)
-\frac{\dot\omega^{(1)}}{\bar f}(\partial_t\bar f+\partial_r\bar f)
-\frac{8\pi G}{\bar f}\Bigl(\bar\Phi_1\dot\Phi_1^{(1)}(\partial_t\bar\omega
+\partial_r\bar\omega)\cr +(\bar\Phi_1\dot\omega^{(1)}+\dot W_1^{(1)})(\partial_t\bar\Phi_1
+\partial_r\bar\Phi_1)+\dot\Phi_1^{(1)}(\partial_t\bar W_1+\partial_r\bar W_1)
-g\bar\Phi_1\dot W_2^{(1)}(\bar A_1+\bar A_2)\Bigr)
\end{eqnarray}

These perturbation equations are linear differential equations of first order in $\dot h_{\mu\nu}$
and $\dot B_\mu^a$ and contain in general  the term  $\ddot k_{\mu\nu}$ and $\ddot C_\mu^a$
In our special case, we see appear in Eq. (38) a  second order term $\ddot k_{\mu\nu}$ and in
Eq. (34) a term $\ddot A_i^{(2)}$.
The remaining perturbation
equations will also contain second order term like $\ddot W_1^{(2)}$.

For the background variables $\bar\Phi_1, \bar W_1$ and $\bar W_2$ we use Eq.(A13).
The resulting equations are similar to the equations found in section 2 for $\Phi_2=0$.
From Eq. (A17) we obtain equations for the background variables $\bar\Omega , \bar f$,
and $\bar\omega$ with back-reaction terms, for example,
\begin{eqnarray}
\partial_t^2\bar\Omega -\partial_r^2\bar\Omega = -\frac{1}{r}\partial_r\bar\Omega
+\frac{1}{\bar\Omega}\Bigl((\partial_t\bar\Omega)^2-(\partial_r\bar\Omega)^2\Bigr)
+\frac{\bar\Omega(\partial_r\bar f)^2}{2\bar f^2}-\frac{\bar\Omega\partial_r\bar f}{\bar fr}
+\frac{\bar f^2\bar\Omega(\partial_r\bar\omega)^2}{2r^2} \cr +\frac{\bar\Omega}{2\tau \bar f^2r^2}
\int[\bar f^4(\dot\omega^{(1)})^2+r^2(\dot f^{(1)})^2]d\xi
+\frac{8\pi G\bar\Omega}{\bar fr^2\tau}\int\Bigl[\bar f^2(\dot W_1^{(1)}+\bar\Phi_1
\dot\omega^{(1)})^2+r^2(\dot\Phi_1^{(1)})^2+\bar f^2(\dot W_2^{(1)})^2\Bigr]d\xi \cr
+\frac{4\pi G\bar\Omega \bar f}{r^2}\Bigl[(\partial_t\bar W_1+\bar\Phi_1\partial_t\bar\omega)^2
+(\partial_r\bar W_1+\bar\Phi_1\partial_r\bar\omega)^2+\frac{r^2}{\bar f^2}(
\partial_r\bar\Phi_1^2+\partial_t\bar\Phi_1^2)\Bigr].
\end{eqnarray}

So we see that a first order HF gravitational and Yang-Mills wave emitted from the
string, will change the metric background component $\bar\Omega$, just as in
the vacuum situation for pure gravitational waves. Moreover, there will be a distortion of the
shape of the wave during its propagation, which does not occur in the pure Einstein case.
In general, this is due to the fact that second order terms appear in the perturbation equations of
$\dot h_{\mu\nu}$ and $\dot B_\mu ^a$.
Now let us consider, for example, the equation for $\dot W_2^{(1)}$, Eq. (36). For $A_1=-A_2$, we can solve
$\dot W_2^{(1)}$ for $\bar f=e^{2p}$. For example for $p=1/4$ we have
\begin{equation}
\dot W_2^{(1)}=\sqrt{r}B(t-r)e^{-2/3r^{1.5}},
\end{equation}
which approaches to zero for large r. Here is $B(t-r)$ an arbitrary function of $(t-r)$.
Together with Eq.(31) we then have to first order
\begin{eqnarray}
\dot W_2^{(1)}=1-1/g+\sqrt{r}B(t-r)e^{-2/3r^{1.5}}+...,
\end{eqnarray}
which is bounded for large r. One can proceed in the same way for the other first order
perturbations by imposing some trial solutions for the background variables~\cite{Slag398}

\section{Conclusions}

Non-Abelian gravity-coupled solitons and black holes could have
played an important role in gauge theories of elementary particles in the early stage of
the universe. Besides the well-known spherically symmetric solutions, axially symmetric
solutions are of interest due to the fact that they admit cosmic strings, which could have
played the role of 'seeds' for the large-scale structure of clusters in the universe.
It is quite evident that the solution found here is a radiating one and there is an
interaction of the solitonlike gravitational waves and the Yang-Mills waves with the
rotating string. This can be seen from figure 5 where the angle deficit  and $\Omega$
is significantly changed by the pulses. This can also be seen from Eq. (41), where the
term on the right-hand side represents the back-reaction of the waves on $\Omega$.
The wave-like solutions found in this model is significantly different from the electrovac
solution found in the Einstein-Maxwell model.
The intervening electro-magnetic waves did not contribute to the angle
deficit asymptotically, while in our model the magnetic components $W_i$ has an impact on the
angle deficit for large r (compare figures 1 and 3). In particularly the number of times the
$W_i$ crosses the r-axis influences the gravitational wave puls.
Further, for a different mass per unit length of the string, a different behavior
of the solution is obtained (compare figures 3 and 4).

In order to conclude if the solution is topological stable, we must solve the perturbation
equations to higher order.
Moreover, a combination of the results of a throughout numerical investigation with the
multiple-scale method pushed to higher orders, will lead to further understanding
of the rich structure of the EYM system on a rotating space time. Specially the question
if a electrically and magnetically charged solution ('dyon') of the rotating cosmic string
exists.
Further, bursts of high-frequency gravitational waves might be detectable by the planned
gravitational wave detectors LIGO/VIRGO and LISA, even for strings at GUT scale~\cite{dam00}.

The issues mentioned above are currently under study by the author.

\appendix

\section{ The EYM equations in the multiple scale formulation}

Let us consider a manifold ${\bf {\cal M}}$ with two different scales, i.e., a mapping
from ${\bf {\cal M}}\times {\cal R}$ into the space of metrics on
${\large{\cal M}}$:
\begin{equation}
\def\x{{\bf x}}
(\x,\xi)\Longrightarrow g(\x,\xi),\qquad \x\in{\cal M}, \xi \in {\cal R}.
\end{equation}
We set $\xi\equiv \omega \Pi({\bf x})$, with $\Pi$ a phase function on ${\cal M}$
of dimension of $length$ and $\omega$ a large parameter of dimension
$(length)^{-1}$.

The multiple-scale method assumes that the metric $g_{\mu\nu}$ and the
YM-potentials $A_\mu^a$ can be written as
\begin{equation}
g_{\mu\nu}=\bar g_{\mu\nu}+\frac{1}{\omega}h_{\mu\nu}(x^\sigma;\xi)+
\frac{1}{\omega^2}k_{\mu\nu}(x^\sigma;\xi)+... ,
\end{equation}
\begin{equation}
A_\mu^a=\bar A_\mu^a +\frac{1}{\omega}B_\mu^a(x^\sigma;\xi)+\frac{1}{\omega^2}
C_\mu^a(x^\sigma;\xi)+... ,
\end{equation}
where $\xi\equiv \omega\Pi(x^\sigma )$ and $\Pi$ a phase function.
The parameter $\omega$ measures the ratio of the fast scale to the slow one.
The rapid variation only occur in the direction of the vector
$l_\sigma \equiv \frac{\partial \Pi}{\partial x^\sigma}$. For a function
$\Psi(x^\sigma ;\xi )$ one has
\begin{equation}
\frac{\partial\Psi}{\partial x^\sigma}= \partial_\sigma \Psi
+\omega l_\sigma \dot \Psi,
\end{equation}
where $\partial_\sigma\Psi \equiv \frac{\partial \Psi}{\partial x^\sigma}
\vert_{\xi fixed}$ and $\dot\Psi \equiv \frac{\partial \Psi}{\partial \xi}\vert
_{x^\sigma fixed}$.
We consider here the case where the magnitude of the perturbation $h_{\mu\nu}$
of the metric with respect to the background $\bar g_{\mu\nu}$ is of order
$\omega^{-1}$ (for different possibilities, such as for example with the leading
term in the metric expansion of order $\omega^{-2}$, see~\cite{taub80}).
Further, we will consider here the hypersurfaces $\Pi(x^\sigma )=cst $ as wave fronts
for the background metric $\bar g$, so (Eikonal equation)
\begin{equation}
l_\alpha l_\beta \bar g^{\alpha\beta}=0.
\end{equation}

Substituting the expansions of the field variables into the equations and
collecting terms of equal orders of $\omega$, one obtains propagation
equations for $\dot B_\mu^a, \ddot C_\mu^a, \dot h_{\mu\nu}$ and
$\ddot k_{\mu\nu}$ and 'back-reaction'
equations for $\bar g_{\mu\nu}$ and $\bar A_\mu^a$. It will be clear from the
propagation equation that there will be a coupling between the high-frequency
gravitational field and the high-frequency behavior of $A_\mu^a$ when the
singularity will be approached.

First we substitute the expansions Eq. (A2) and (A3) into the
YM-equation Eq. (4). We obtain for the order $\omega$ equation
\begin{equation}
\bar g^{\mu\alpha}\Bigl[l_\mu l_\alpha \ddot B_\nu ^a -l_\nu \l_\alpha
\ddot B_ \mu ^a \Bigr] =0.
\end{equation}
For the order $\omega^0$-equation we obtain
\begin{eqnarray}
&&\bar g^{\mu\alpha} \Bigl[\bar\nabla_\alpha\bar {\cal F}_{\mu\nu}^a+
\Upsilon_{\alpha\mu}^\lambda \bar{\cal F}_{\nu\lambda}^a
- \Upsilon_{\alpha\nu}^\lambda \bar{\cal F}_{\mu\lambda}^a
+\dot B_\nu^a\Bigl(l_{\mu ,\alpha}-(\bar\Gamma_{\alpha\mu}^\lambda
+\Upsilon_{\alpha\mu}^\lambda )l_\lambda \Bigr)-\dot B_\mu^a\Bigl(l_{\nu ,\alpha} -
(\bar\Gamma_{\alpha\nu}^\lambda +\Upsilon_{\alpha\nu}^\lambda )l_\lambda\Bigr)\cr
&& +l_\mu \Bigl(\dot B_{\nu ,\alpha}-(\bar\Gamma_{\alpha\nu}^\lambda +
\Upsilon_{\alpha\nu}^\lambda )\dot B_\lambda^a\Bigr)
-l_\nu\Bigl(\dot B_{\mu ,\alpha}^a-(\bar\Gamma_{\alpha\mu}^\lambda
+\Upsilon_{\alpha\mu}^\lambda )\dot B_\lambda^a\Bigr)
+l_\alpha (\dot B_{\nu ,\mu}^a-\dot B_{\mu ,\nu}^a)\cr
&& \qquad +l_\alpha (l_\mu \ddot C_\nu^a -l_\nu \ddot C_\mu^a)+g\epsilon^{abc}\Bigl(
l_\alpha (\bar A_\mu^b \dot B_\nu^c+\bar A_\nu^c \dot B_\mu^b)
+\bar A_\alpha^b (\bar {\cal F}_{\mu\nu}^c+l_\mu\dot B_\nu^c
-l_\nu\dot B_\mu^c )\Bigr)\Bigr]\cr  && \qquad\qquad\qquad -h^{\mu\alpha} l_\alpha
(l_\mu\ddot B_\nu^a -l_\nu\ddot B_\mu^a)=0,
\end{eqnarray}
with $\Upsilon_{\mu\nu}^\lambda \equiv\frac{1}{2}\bar g^{\sigma\lambda}(l_\mu\dot h_{\nu\sigma}-
l_\nu\dot h_{\mu\sigma}-l_\sigma\dot h_{\mu\nu}).$

Substituting the expansions into the Einstein equations Eq. (3), we obtain
for the order $\omega$ equation

\begin{equation}
R_{\mu\nu}^{(-1)}=l_\nu\dot\Upsilon_{\mu\sigma}^\sigma -
l_\sigma\dot\Upsilon_{\mu\nu}^\sigma =0.
\end{equation}
For the $\omega^{0}$-equation we obtain
\begin{eqnarray}
\bar R_{\mu\nu}+R_{\mu\nu}^{(0)}=-8\pi G\Bigl\{\bar g^{\lambda\beta}(
\bar{\cal F}_{\mu\lambda}^a +l_\mu\dot B_\lambda ^a-l_\lambda\dot B_\mu^a)
(\bar{\cal F}_{\nu\beta}^a+l_\nu\dot B_\beta^a -l_\beta\dot B_\nu^a)
\cr +\frac{1}{4}\bar g_{\mu\nu}\bar g^{\sigma\alpha}\bar g^{\tau\beta}
(\bar{\cal F}_{\alpha\beta}^a+l_\alpha\dot B_\beta^a -l_\beta\dot B_\alpha^a)
(\bar{\cal F}_{\sigma\tau}^a+l_\sigma\dot B_\tau^a-l_\tau\dot B_\sigma^a)\Bigr\},
\end{eqnarray}
with $\bar R_{\mu\nu}$ the background Ricci tensor and $R_{\mu\nu}^{(0)}$ an
expression in $\ddot k, h\ddot h, \dot h^2$ and $\dot h$ (see~\cite{choq169}).
We can simplify the equations considerable. If we use Eq. (A5), then Eq. (A6)
becomes
\begin{equation}
l^\mu \ddot B_\mu^a =0,
\end{equation}
so for periodic $B_\mu^a$ we have
\begin{equation}
l^\mu B_\mu^a=0.
\end{equation}
Using Eq. (A8) and Eq. (A10)  we obtain from Eq. (A7)
\begin{eqnarray}
\bar g^{\mu\alpha}\Bigl[\bar\nabla_\alpha\bar{\cal F}_{\mu\nu}^a-
\Upsilon_{\alpha\nu}^\lambda\bar{\cal F}_{\mu\lambda}^a +\dot B_\nu^a
\bar\nabla_\alpha l_\mu -\dot B_\mu^a\bar\nabla_\alpha l_\nu
+l_\mu\bar\nabla_\alpha\dot B_\nu^a-l_\nu\bar\nabla_\alpha\dot B_\mu^a
+l_\alpha(\dot B_{\nu,\mu}^a -\dot B_{\mu,\nu}^a )\Bigr] \cr
+l_\nu l_\alpha h^{\mu\alpha} \ddot B_\mu ^a
-l_\nu l^\mu \ddot C_\mu^a +g\epsilon^{abc}\bar g^{\mu\alpha}\Bigl[
\bar A_\alpha^b\bar{\cal F}_{\mu\nu}^c +2 l_\alpha \bar A_\mu^b\dot B_\nu^c
-l_\nu\bar A_\alpha^b\dot B_\mu^c\Bigr]=0,
\end{eqnarray}
with $\bar\nabla$ the covariant derivative with respect to the
background metric $\bar g_{\mu\nu}$.
Integrating this equation with respect to $\xi$ yields
\begin{equation}
\bar{\cal D}^\mu\bar{\cal F}_{\mu\nu}^a=0.
\end{equation}
Substituting back this equation into Eq. (A12) we obtain the
propagation equation for the YM-field
\begin{eqnarray}
\bar\nabla^\mu (l_\mu\dot B_\nu^a-l_\nu \dot B_\mu^a)+l^\mu (\bar\nabla_\mu
\dot B_\nu^a -\bar\nabla_\nu\dot B_\mu^a)
-\bar g^{\mu\alpha}\Upsilon_{\alpha\nu}^\lambda\bar {\cal F}_{\mu\lambda}^a
+l^\alpha l_\nu h_\alpha^\lambda\ddot B_\lambda^a \cr
-l_\nu l^\mu \ddot C_\mu^a
+g\epsilon^{abc}\bar g^{\mu\alpha}(2l_\alpha\bar A_\mu^b\dot B_\nu^c-
l_\nu\bar A_\alpha^b\dot B_\mu^c )=0.
\end{eqnarray}
Multiplying this propagation equation with $\dot B^{\nu a}$,
we obtain the 'conservation'-equation
\begin{equation}
\bar\nabla_\alpha (l^\alpha \dot B_\nu^a\dot B^{\nu a} )=l^\mu\dot h_\nu^\lambda
\dot B^{\nu a}\bar{\cal F}_{\mu\lambda}^a-2g\epsilon^{abc}l^\mu
\dot B_\nu^c\dot B^{\nu a}\bar A_\mu^b,
\end{equation}
so $\dot B_\nu^a\dot B^{\nu a}$ is not conserved, unless the right-hand side
is zero.
Multiplying Eq. (A14)  with $l^\nu$, we obtain
\begin{equation}
l^\nu (l^\lambda\dot h_\nu^\mu -l^\mu\dot h_\nu^\lambda )
\bar{\cal F}_{\mu\lambda}^a =0.
\end{equation}
For the choice $l^\nu h_{\nu\mu}=0 $ and hence from Eq. (A8), $h=0$, the
equations simplify again.\\
Integrating Eq. (A9) with respect to $\xi$, we obtain
\begin{equation}
\bar R_{\mu\nu}=-\frac{1}{4\tau} l_\mu l_\nu \int(\dot h_\rho^\sigma
\dot h_\sigma^\rho -\frac{1}{2}\dot h^2) d\xi -\frac{8\pi G}{\tau} l_\mu l_\nu
\int\dot B^{\alpha a}\dot B_\alpha ^a d\xi -8\pi G\bar{\cal T}_{\mu\nu},
\end{equation}
with $\tau$ the period of high-frequency components. In the right-hand side
we have the back-reaction term due to the high-frequency gravitational
perturbation, the high-frequency YM-perturbations and the background energy-
momentum term $\bar{\cal T}$.
Substitution back Eq. (A17) into Eq. (A9), we obtain the perturbation equations for
$h_{\mu\nu}$
\begin{eqnarray}
R_{\mu\nu}^{(0)}=l_\mu l_\nu\Bigl[
\frac{1}{4\tau}\int(\dot h_\rho^\sigma\dot h_\sigma^\rho-\frac{1}{2}\dot h^2) d\xi
+\frac{8\pi G}{\tau}\int\dot B^{\alpha a}\dot B_\alpha^a d\xi\Bigr]
-8\pi G\Bigl[(l_\mu l_\nu \dot B^{\lambda a}\dot B_\lambda^a \cr
+\bar{\cal F}_{
\mu\lambda}^a (l_\nu\dot B^{\lambda a}-l^\lambda\dot B_\nu^a )+
\bar{\cal F}_{\nu\lambda}^a (l_\mu\dot B^{\lambda a}-l^\lambda\dot B_\mu^a )
+\frac{1}{2}\bar g_{\mu\nu}\bar F_{\alpha\beta}(l^\alpha\dot B^{\beta a}-
l^\beta\dot B^{\alpha a})\Bigr]
\end{eqnarray}

\newpage

Figure 1. Plot of the metric- and YM components for the initial values:
$\Omega=Ar^{p^2-p}(A=3,p=0.25)$, $f=r^{2p}$, $\omega=\frac{3.5r}{r^2+1}$
$\Phi_1=1-e^{-0.8r}$, $w=0.8-e^{-(r-2)^2}$,$\Phi_2=e^{-0.32r}-1$.G=4.5  g=1.25. \\

Figure 2. As figure 1, with  p=0.1, G=0.6 \\

Figure 3. As figure 2, for the initial $W_1$: $W_1=-\tanh (r-1)$\\

Figure 4. As figure 3, for a smaller value of the mass per unit length of the string:
p=0.09 in stead of 0.25.                                          \\

Figure 5. A long time run for $\Phi_1=1-e^{-0.5r}$

\newpage
\input epsf
\centerline
{\hfil\epsfysize=45mm\epsfbox{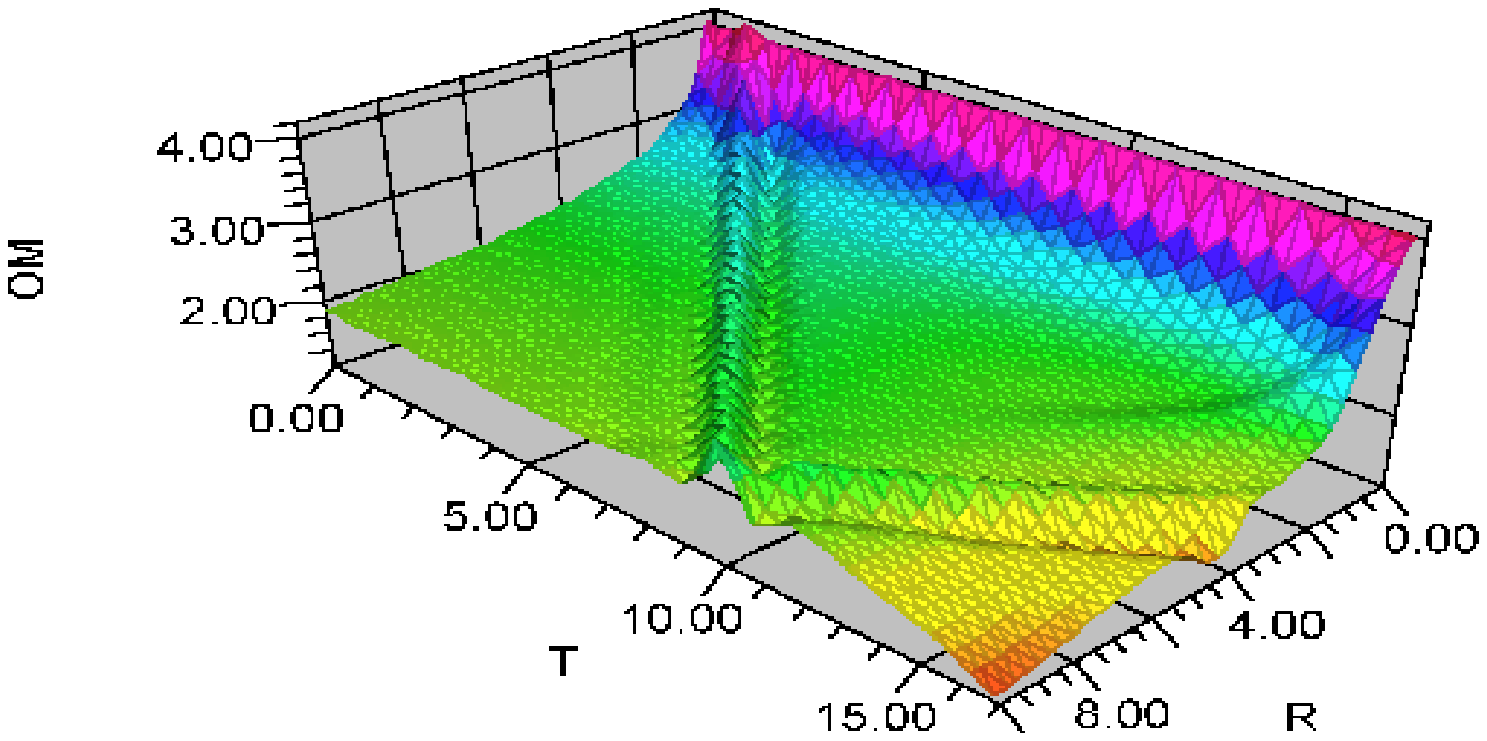}
\hfil\epsfysize=45mm\epsfbox{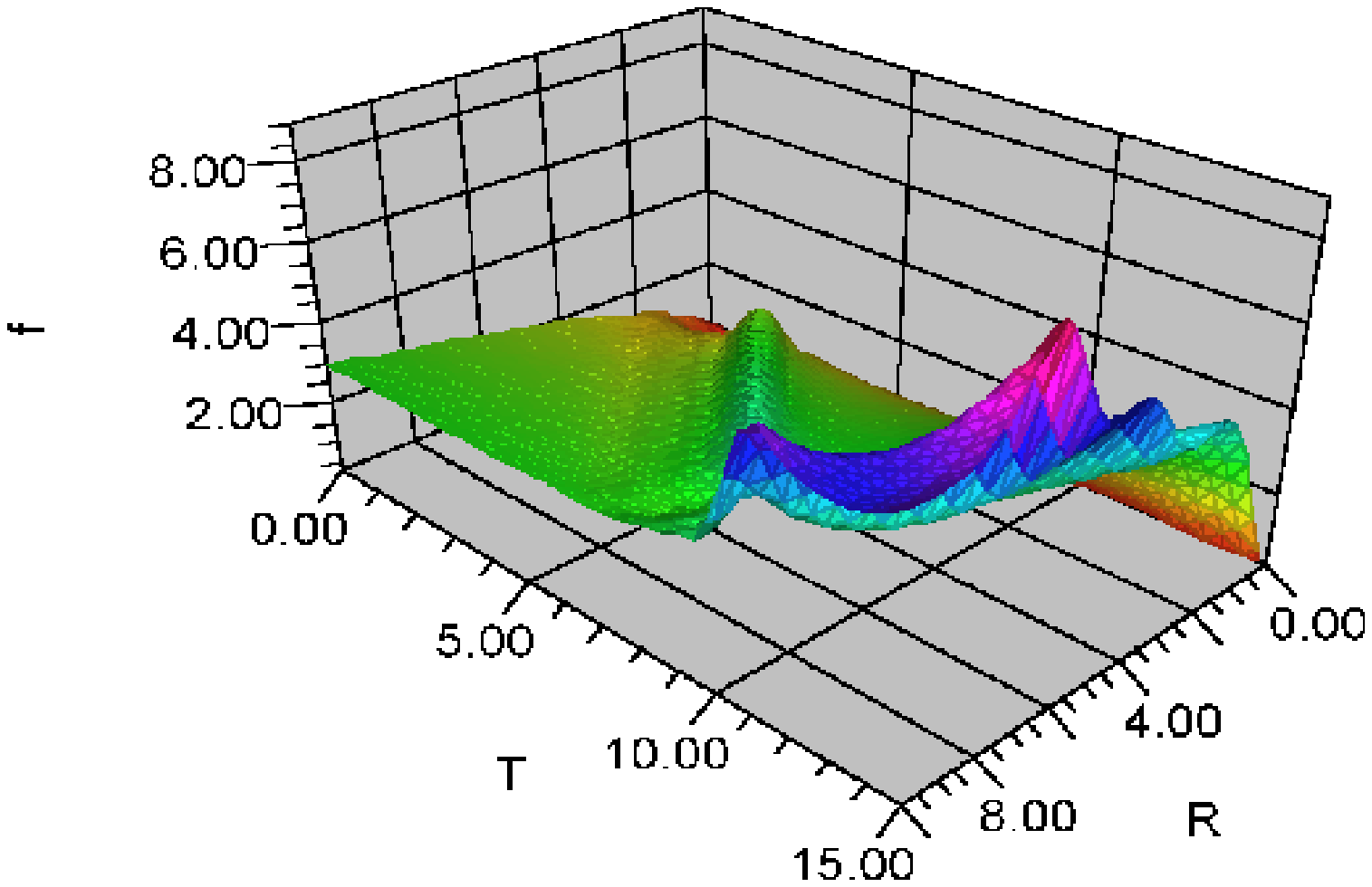}
\hfil}
\centerline
{\hfil\epsfysize=45mm\epsfbox{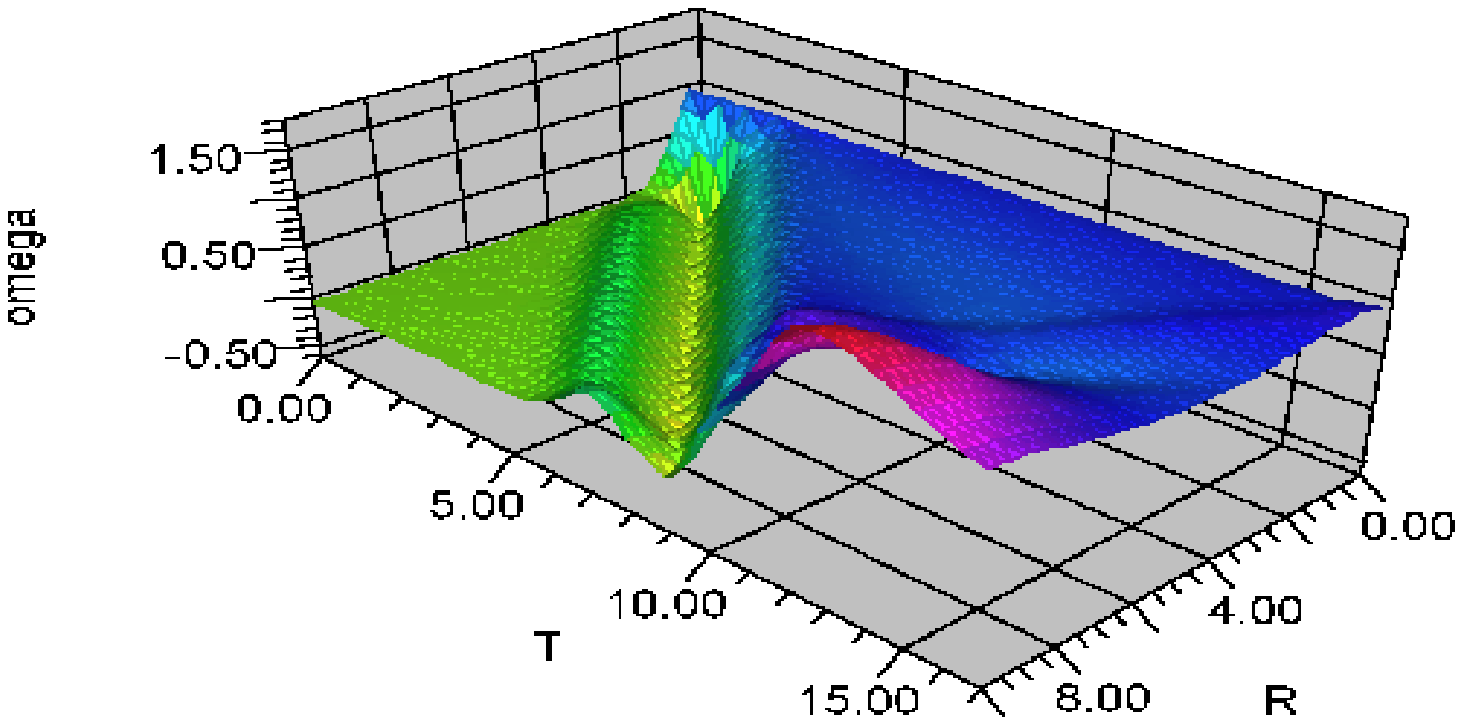}
\hfil\epsfysize=45mm\epsfbox{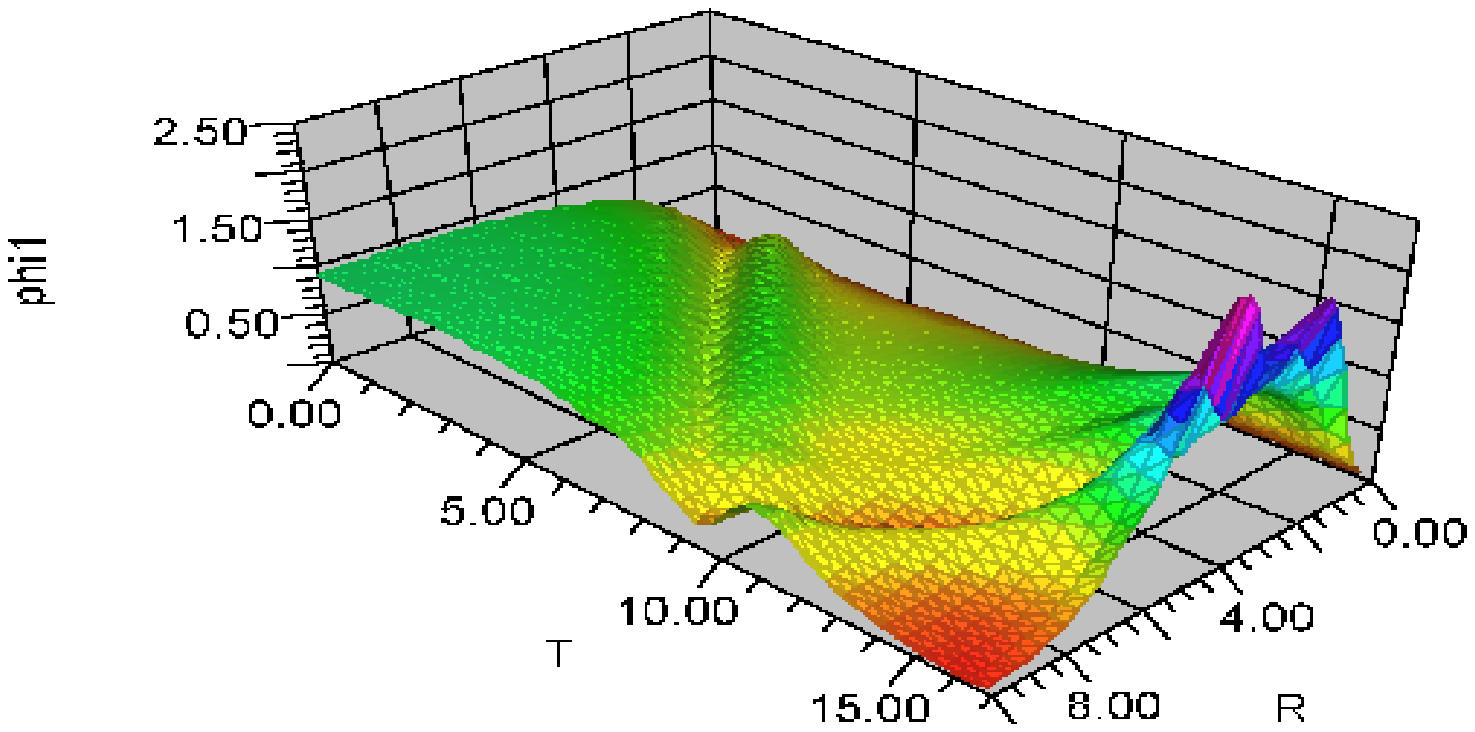}
\hfil}
\centerline
{\hfil\epsfysize=45mm\epsfbox{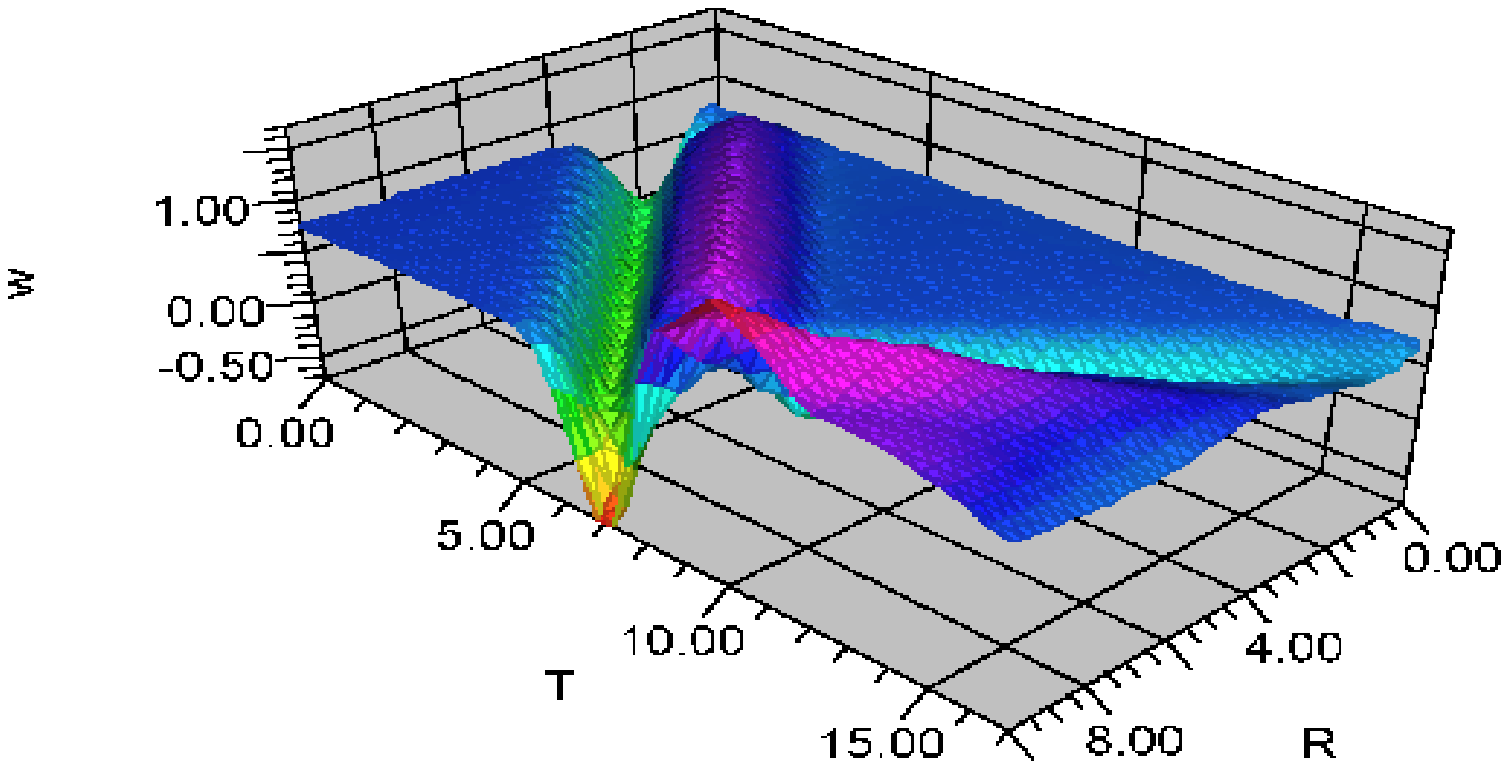}
\hfil\epsfysize=45mm\epsfbox{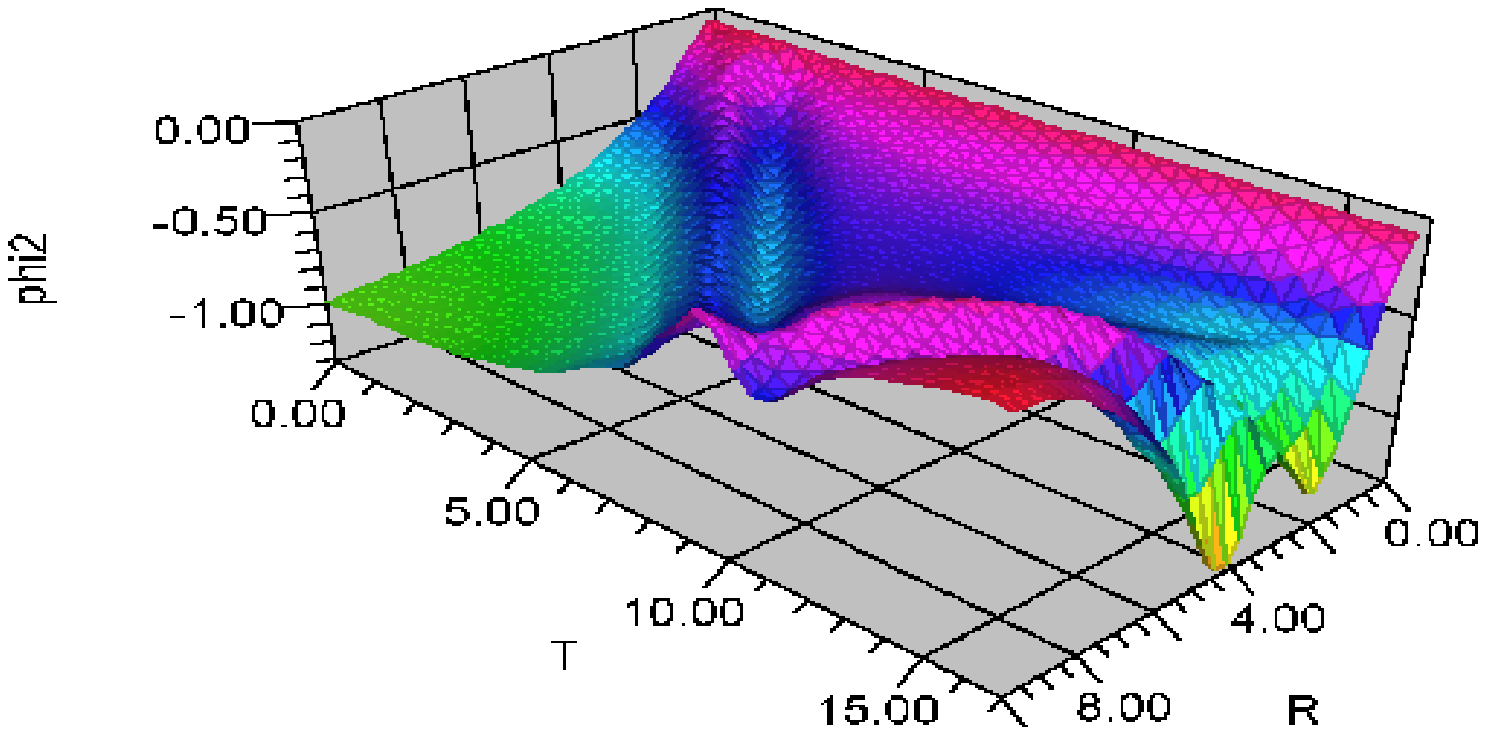}
\hfil}
\centerline
{\hfil\epsfysize=45mm\epsfbox{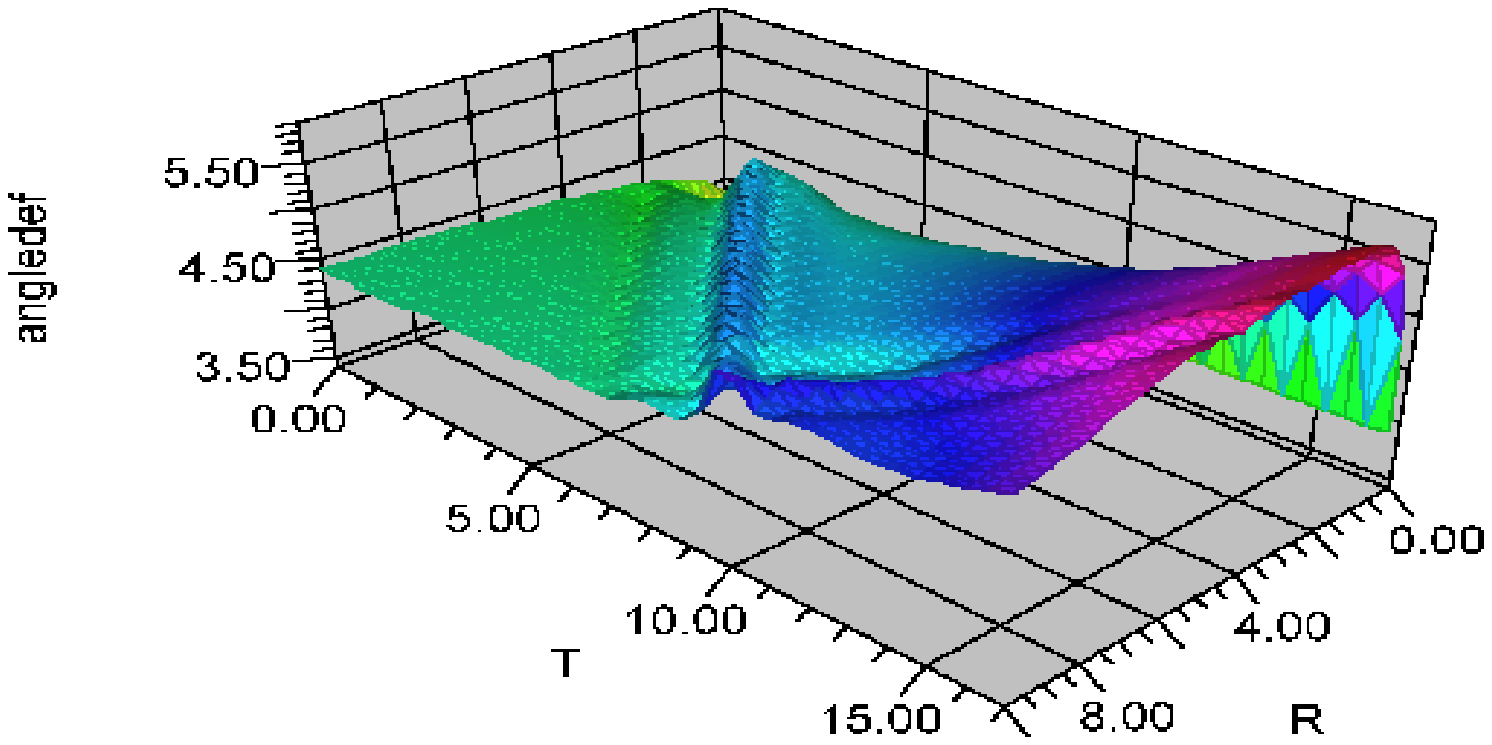}
\hfil\epsfysize=45mm\epsfbox{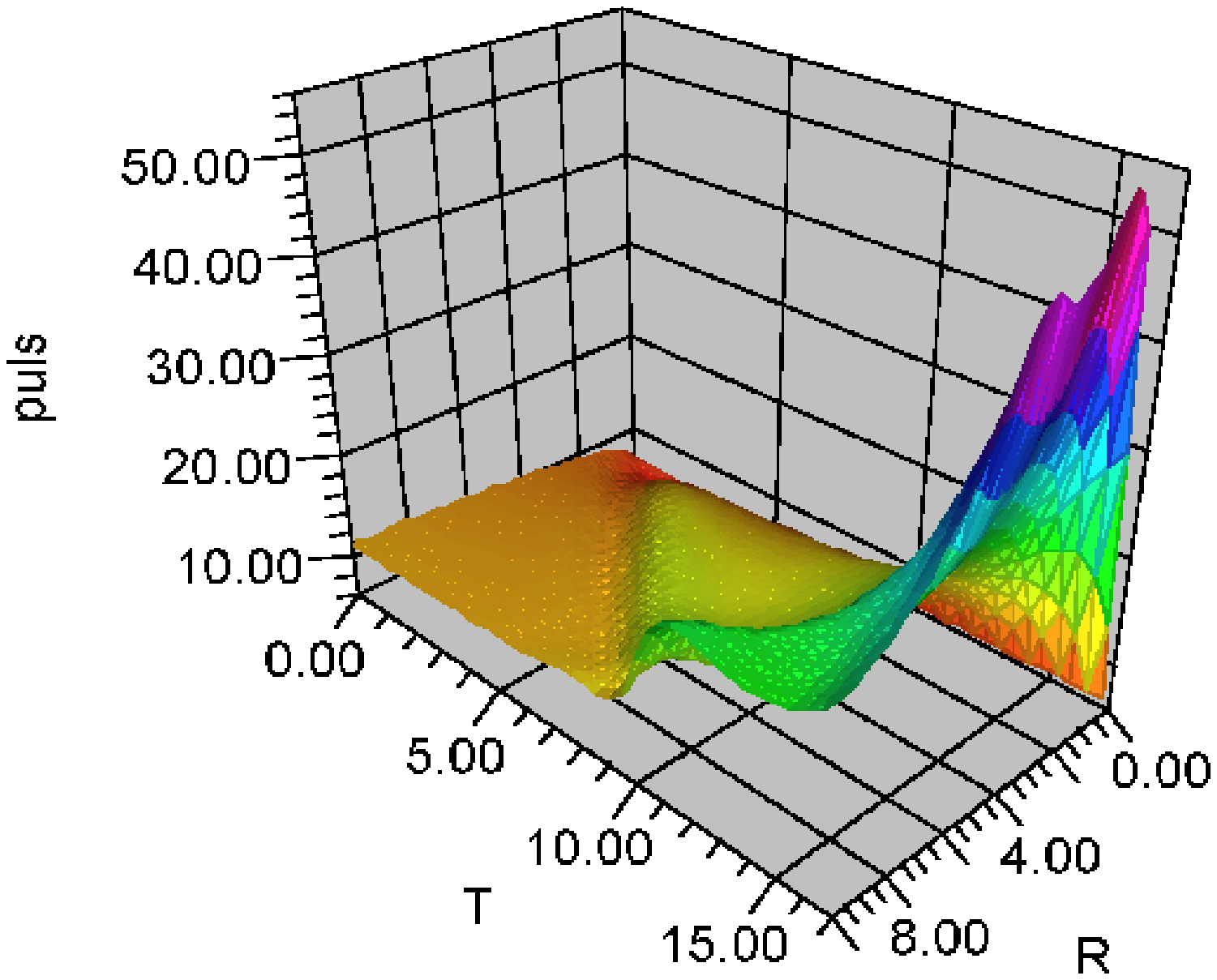}
\hfil}
\centerline
{\hfil\epsfysize=45mm\epsfbox{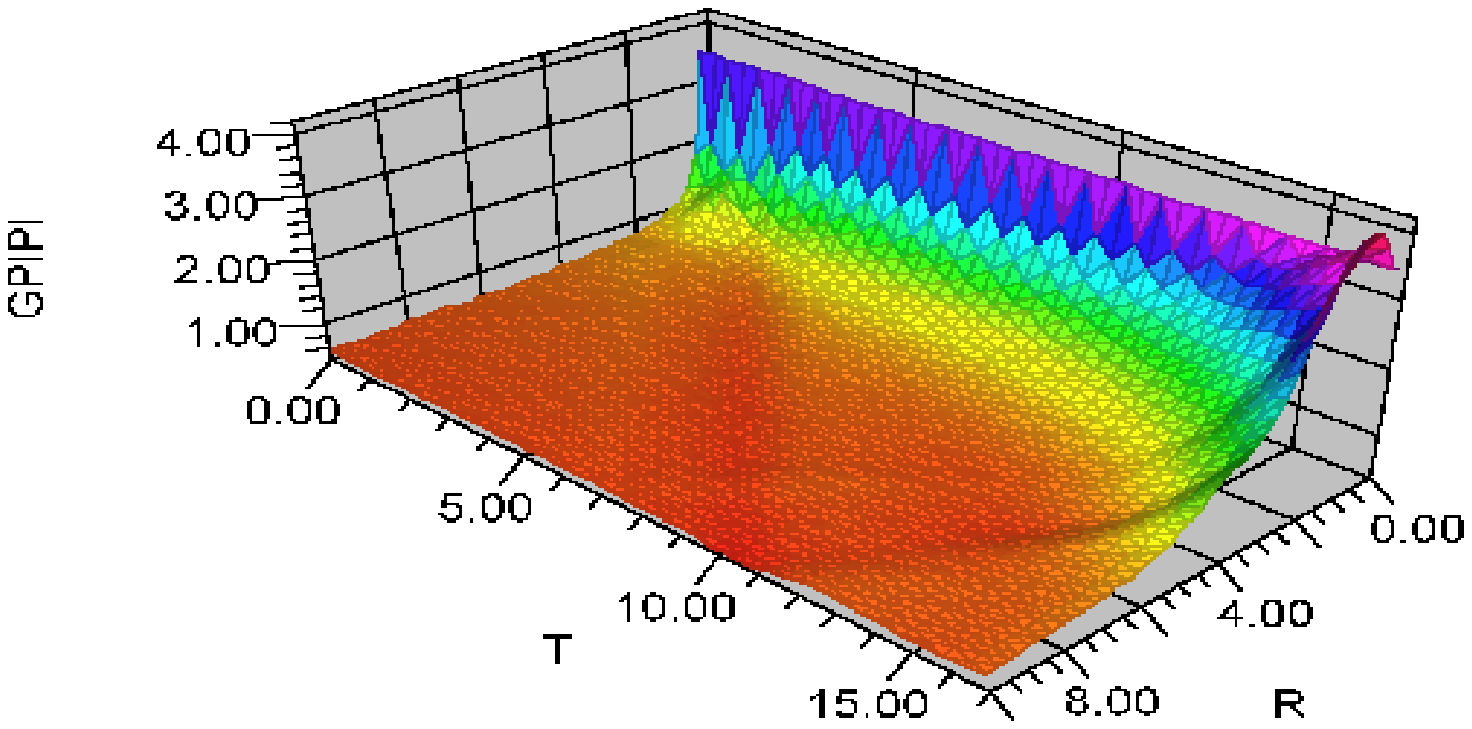}
 \hfil\epsfysize=45mm\epsfbox{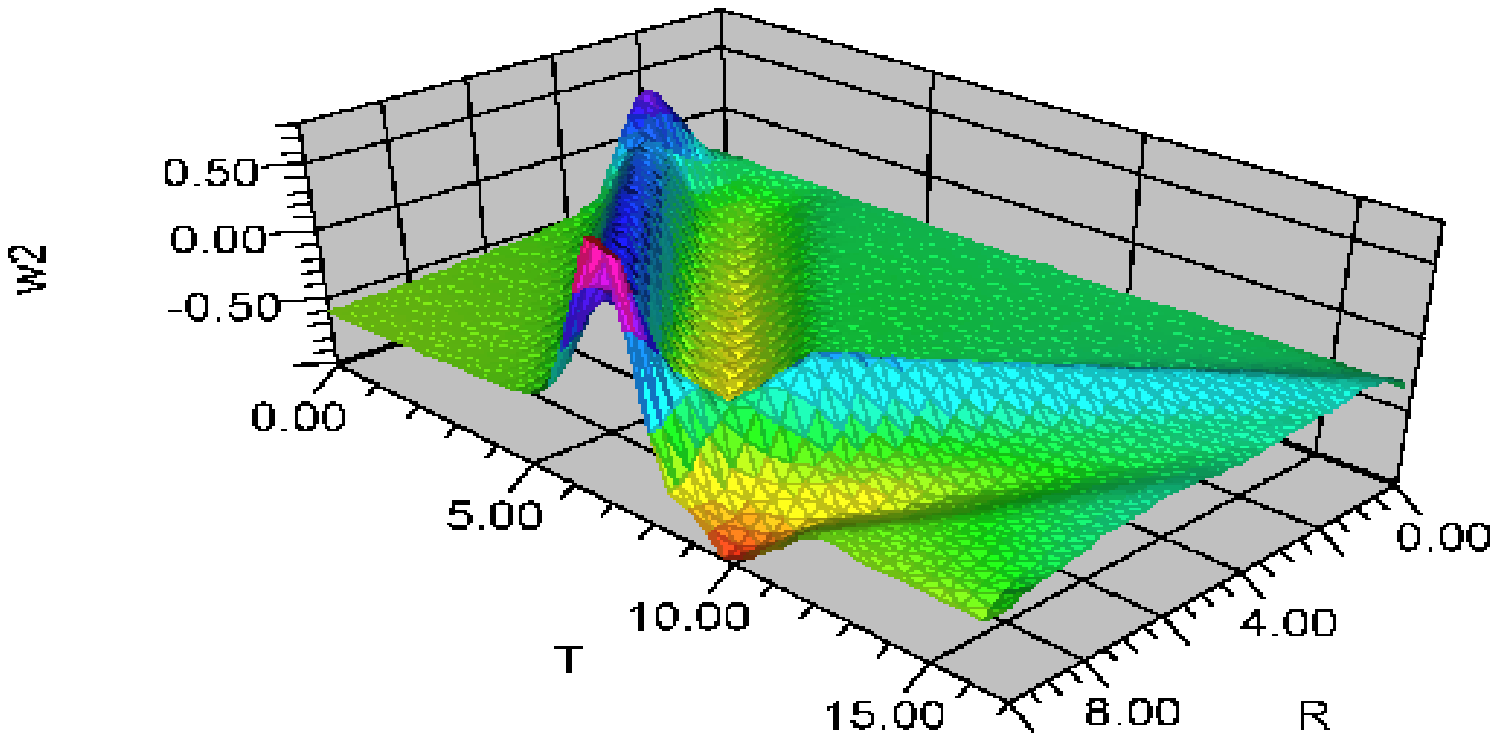}
\hfil}

\end{document}